# A STATISTICAL ANALYSIS OF MEMORY CD8 T CELL DIFFERENTIATION: AN APPLICATION OF A HIERARCHICAL STATE SPACE MODEL TO A SHORT TIME COURSE MICROARRAY EXPERIMENT[1]

By Haiyan Wu, Ming Yuan, Susan M. Kaech
and M. Elizabeth Halloran

*Emory University, Georgia Institute of Technology, Yale University
and University of Washington*

CD8 T cells are specialized immune cells that play an important role in the regulation of antiviral immune response and the generation of protective immunity. In this paper we investigate the differentiation of memory CD8 T cells in the immune response using a short time course microarray experiment. Structurally, this experiment is similar to many in that it involves measurements taken on independent samples, in one biological group, at a small number of irregularly spaced time points, and exhibiting patterns of temporal nonstationarity. To analyze this CD8 T-cell experiment, we develop a hierarchical state space model so that we can: (1) detect temporally differentially expressed genes, (2) identify the direction of successive changes over time, and (3) assess the magnitude of successive changes over time. We incorporate hidden Markov models into our model to utilize the information embedded in the time series and set up the proposed hierarchical state space model in an empirical Bayes framework to utilize the population information from the large-scale data. Analysis of the CD8 T-cell experiment using the proposed model results in biologically meaningful findings. Temporal patterns involved in the differentiation of memory CD8 T cells are summarized separately and performance of the proposed model is illustrated in a simulation study.

## 1. Introduction.

### 1.1. *Background.*
Time course microarray experiments have become increasingly popular in the study of dynamic biological processes due to their

---

Received November 2006; revised April 2007.

[1]Supported in part by NIH Grant 5 U19-AI057266-04.

ability to monitor tens of thousands of genes over time. As of June 2004 [Ernst, Nau and Bar-Joseph (2005)], 80% of time course microarray experiments were short time series with fewer than eight time points, according to the Stanford Microarray Database [Gollub et al. (2003)]. In this paper we are particularly interested in a short time course microarray experiment on memory CD8 T cell differentiation originally described and analyzed in Kaech et al. (2002a). This CD8 T-cell experiment was done in the context of a large research effort to understand immune memory in Rafi Ahmed's laboratory of the Emory Vaccine Research Center. Here immune memory refers to the ability of the immune system to remember its first exposure to a specific antigen and to mount a rapid and aggressive response to a second exposure. In the immune system, CD8 T cells are specialized immune cells that play an important role in the regulation of antiviral response and the generation of protective immunity. In response to a viral infection, naïve CD8 T cells differentiate into effector CD8 T cells that control the infection and the effector CD8 T cells that survive continue to differentiate into long-lived protective memory CD8 T cells [Kaech, Wherry and Ahmed (2002b)].

In this CD8 T-cell experiment, acute lymphocytic choriomeningitis virus Armstrong (LCMV) infection of mice was used as a model system to study memory CD8 T cell differentiation. Genetically identical, uninfected mice were sacrificed on the baseline day (naïve) to obtain naïve CD8 T cells. Other genetically identical mice were infected with LCMV on the baseline day. Then mice were sacrificed at day 8 (d8) and day 15 (d15) to obtain effector CD8 T cells, and at greater than day 30 (Imm) to obtain memory CD8 T cells. Affymetrix MG-U74AV2 arrays were used to measure 12,488 genes in P14 CD8 T cells from mouse spleens at these four time points. For each chip, cells from at least three mice were pooled to obtain sufficient RNA for MG-U74AV2 hybridization. Structurally, this CD8 T-cell experiment is similar to many in that it involves measurements taken on different mice (independent sampling), in one biological group, at a small number of irregularly spaced time points, and exhibiting patterns of temporal nonstationarity. The goal of the analysis is to assist investigators in understanding the underlying system biology by identifying temporally differentially expressed (TDE) genes and characterizing temporal changes involved in memory CD8 T cell differentiation.

In the original analysis, Kaech et al. (2002a) selected genes based on whether their average gene expression levels changed (decreased or increased) at any time point by at least 1.7 fold (original linear scale) compared to the first time point, generating a set of 431 genes. They applied a K-means clustering algorithm on the selected genes and found six major patterns out of 10 clusters. In this original analysis, the temporal aspects of the data were ignored and both the fold change cutoff in the selection method and the number of clusters in K-means clustering were chosen arbitrarily. Although they identified several individual genes known to be important in differentiating naïve, effector and memory CD8 T cells, the biological meaning of the obtained clusters is not clear and the interpretation of clustering results is not straightforward. To improve interpretation, we re-analyzed this CD8 T-cell experiment by focusing on the direction (upregulation, downregulation and no change) and the magnitude of the gene-specific successive differences (changes) in the mean gene expression levels (log base 2 scale) over time.

1.2. *Analysis of time course microarray data.* Methods up to now for time course microarray experiments can be divided into two classes: (1) methods extended from those for static microarray experiments, and (2) methods extended from time series analysis. Examples of the first type of methods include hierarchical clustering [Eisen et al. (1998) and Spellman et al. (1998)], K-means clustering [Tavazoie et al. (1999), Kaech et al. (2002a) and Bar–Joseph et al. (2002)], self-organizing maps [Tamayo et al. (1999)], singular value decomposition [Alter, Brown and Botstein (2000) and Wall, Dyck and Brettin (2001)], ANOVA-based analysis [Park et al. (2003)] and pairwise analysis. Ignoring the temporal aspects of data, these types of methods have the potential to suffer from low sensitivity. Examples of methods inspired by time series analysis include Auto-Regression (AR) based models, multivariate Normal models, B-splines based models, and hidden Markov models (HMMs). Ramoni, Sebastiani and Cohen (2002) represented gene expression sequences as stationary series produced from a finite number of AR processes and applied an agglomerative Bayes clustering algorithm to search gene clusters. Using the multivariate Normal distribution, Tai and Speed (2006) developed a multivariate hierarchical empirical Bayes model to identify TDE genes. Storey et al. (2005) proposed a general framework for time course microarray experiments by modeling a gene-specific expression sequence as a linear expansion of B-spline basis functions. Hong and Li (2006) proposed to add a hierarchical structure into a B-spline based model to identify genes temporally differentially expressed between two biological groups [see Spellman et al. (1998), Klevecz (2000) and Heard, Holmes and Stephens (2006) for other choices for basis functions]. Schliep, Schönhuth and Steinhoff (2003, 2004) proposed to use a mixture of hidden Markov models to model the gene expression level sequence and to cluster genes where a heuristic cluster deletion and splitting procedure is utilized to determine the number of clusters. Yuan and Kendziorski (2006) constructed a hidden Markov model to infer the gene-specific relative relationship of multiple biological groups over time. Zhou and Wakefield (2006) proposed a first-order random walk model to partition TDE genes where the optimal number of partitions is chosen based on posterior probabilities via birth-death MCMC. Although these methods are useful in the analysis of time course microarray experiments, none of them can be used to investigate both the direction and the magnitude of successive changes in the mean gene expression levels in the CD8 T-cell experiment.

To add to these useful methods, we develop a hierarchical state space model using the observed gene expression levels along with the direction and the magnitude of successive changes in the mean gene expression levels. To improve the sensitivity of detecting TDE genes, we incorporate HMMs into our model to utilize the correlation information embedded in the time series [Yuan and Kendziorski (2006)] and set up our model in an empirical Bayes framework to utilize information from the gene population [Efron et al. (2001) and Newton et al. (2001)]. Analysis of the CD8 T-cell experiment using the proposed hierarchical state space model results in the identification of more significant genes (over 1200 vs 431 in the original analysis) and the discovery of new important temporal patterns such as continuous upregulation over time (see Section 3).

## 2. Hierarchical state space model.

2.1. *Hierarchical state space model for the CD8 T-cell experiment.* In the CD8 T-cell experiment, expression levels were measured for $G = 12{,}488$ genes at $T = 4$ time points where large $G$ and small $T$ are typical of many time course microarray studies. Replicates were independently sampled with three replicates at day 15 and four replicates at each of the other three time points. We denote by $x_{gtk}$ the observed gene expression level (log base 2 scale) for gene $g$ at the $t$th time point on array $k$ for $k = 1, \ldots, n_t$ and denote by $\mathbf{x}_{gt} = (x_{gt1}, \ldots, x_{gtn_t})$ all the observed gene expression levels for gene $g$ at the $t$th time point. The observed gene expression levels can be represented by a $12{,}488 \times 15$ matrix whose rows represent genes and columns represent arrays. The typical data layout is shown in Table 1.

Our interest lies in the gene-specific mean expression levels over time, a latent mean sequence denoted by $\boldsymbol{\mu}_g = (\mu_{g1}, \ldots, \mu_{gT})$ where $\mu_{gt} = E(x_{gtk})$.

TABLE 1
*Data layout for microarray experiments with one biological group and independent sampling*

|  | **Time 1** | | | **Time 2** | | | . . . | **Time $T$** | | |
|---|---|---|---|---|---|---|---|---|---|---|
|  | 1 | . . . | $n_1$ | 1 | . . . | $n_2$ | . . . | 1 | . . . | $n_T$ |
| Gene 1 | $x_{111}$ | . . . | $x_{11n_1}$ | $x_{121}$ | . . . | $x_{12n_2}$ | . . . | $x_{1T1}$ | . . . | $x_{1Tn_T}$ |
| Gene 2 | $x_{211}$ | . . . | $x_{21n_1}$ | $x_{221}$ | . . . | $x_{22n_2}$ | . . . | $x_{2T1}$ | . . . | $x_{2Tn_T}$ |
| $\vdots$ | $\vdots$ | $\vdots$ | $\vdots$ | $\vdots$ | $\vdots$ | $\vdots$ | $\vdots$ | $\vdots$ | $\vdots$ | $\vdots$ |
| Gene $G$ | $x_{G11}$ | . . . | $x_{G1n_1}$ | $x_{G21}$ | . . . | $x_{G2n_2}$ | . . . | $x_{GT1}$ | . . . | $x_{GTn_T}$ |

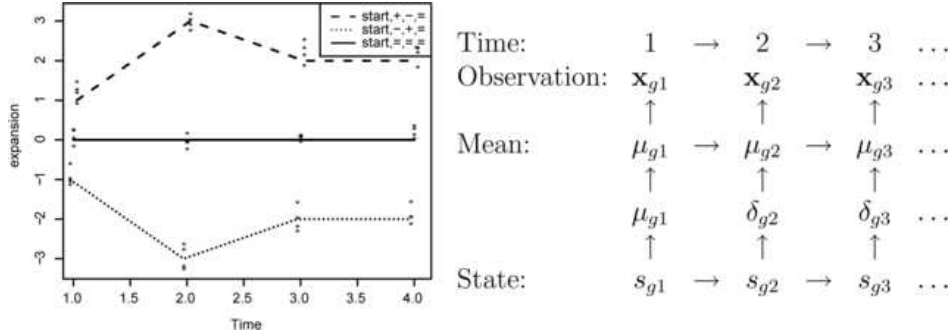

Fig. 1. *An example of three temporal patterns with independent sampling and illustration of the first-order hierarchical state space model: In the left panel of the figure, the y-axis represents the gene expression level (log base 2 scale) and the x-axis represents the time points. Solid lines represent the underlying mean gene expression levels (log base 2 scale) over time and stars represent the observed gene expression levels of three genes in an experiment with four replicates at each time point.*

Specifically, we are interested in the magnitude of successive changes in $\mu_g$, a latent sequence composed by $\delta_{gt} = \mu_{gt} - \mu_{g(t-1)}$ for $t = 2, \ldots, T$, and the direction of successive changes in $\mu_g$, a latent state sequence denoted by $\mathbf{s}_g = (s_{g1}, \ldots, s_{gT})$. Here the state at the first time point is defined as "start" and the state at each of the other time points could be upregulation ($+ : \delta_{gt} > 0$), downregulation ($- : \delta_{gt} < 0$), or no change ($= : \delta_{gt} = 0$). This definition leads to a total of $3^{T-1}$ possible temporal patterns for an experiment with $T$ time points (denote by $\mathcal{S}$) and thus a total of 27 possible temporal patterns for the CD8 T-cell experiment. The left panel of Figure 1 shows an example with three temporal patterns in an experiment with 4 time points. The goal of detecting TDE genes is equivalent to finding genes with either the upregulation or downregulation state at any time point after the first time point. The goal of characterizing TDE genes can be achieved by finding all distinctive temporal patterns.

For the CD8 T-cell experiment, we develop a hierarchical state space model in three levels (the observation level, the mean level and the state level) as follows. In the observation level, the observed gene expression levels at time $t$ ($\mathbf{x}_{gt}$) are regarded as a random sample independently produced from a Normal distribution with mean $\mu_{gt}$ and a common standard deviation $\sigma$, similar to Kendziorski et al. (2003). In the mean level, the successive change of the mean gene expression level under the state of no change is 0 by definition and the successive change of the mean gene expression level under the upregulation (downregulation) state is regarded as a random sample independently produced from an unknown 0-left-truncated (0-right-truncated) Normal distribution. In the state level, the state sequence is regarded as a random sample independently produced from an unknown discrete Markov process in which various orders of Markov dependence are allowed, similar to Yuan and Kendziorski (2006). The order of the proposed hierarchical state space model is defined by the order of the Markov process in the state level. See the right panel of Figure 1 for an illustration of the first-order hierarchical state space model. The proposed model is established in an empirical Bayes framework to utilize information from the large gene population. The marginal distribution of $\mathbf{x}_g = (\mathbf{x}_{g1}, \ldots, \mathbf{x}_{gT})$ in such case is

$$f(\mathbf{x}_g) = \sum_{\mathbf{v} \in \mathcal{S}} \Pr(\mathbf{s}_g = \mathbf{v}) f(\mathbf{x}_g | \mathbf{s}_g = \mathbf{v}),$$

a mixture of the conditional distributions of $\mathbf{x}_g$ given state series $\mathbf{v}$. Here $\Pr(\mathbf{s}_g = \mathbf{v}) = \Pr(\mathbf{v})$ is the population proportion of the state series $\mathbf{v}$ with the constraint $\sum_{\mathbf{v} \in \mathcal{S}} \Pr(\mathbf{v}) = 1$; $f(\mathbf{x}_g | \mathbf{s}_g = \mathbf{v}) = \int f(\mathbf{x}_g | \boldsymbol{\mu}_g) f(\boldsymbol{\mu}_g | \mathbf{s}_g = \mathbf{v}) \, d\boldsymbol{\mu}_g$ is the conditional distribution of $\mathbf{x}_g$ given state series $\mathbf{v}$. To simplify notation, $f(\cdot | \cdot)$ and $f(\cdot)$ are used throughout to denote a generic conditional and marginal density function respectively. Likewise, $\Pr(\cdot | \cdot)$ and $\Pr(\cdot)$ are used throughout to denote a generic conditional and marginal probability function respectively.

To fit the proposed hierarchical state space model, we need to estimate parameters specifying the Normal distribution in the observation level, parameters specifying the truncated Normal distributions in the mean level, and parameters specifying the Markov process in the state level where all these parameters (denoted by $\boldsymbol{\theta}$) are universal for all genes. Model fitting proceeds by an implementation of the EM algorithm [Dempster, Laird and Rubin (1977)] to produce estimates of fixed effects $\boldsymbol{\theta}$ and posterior distributions for latent mean sequences $\boldsymbol{\mu}_g$ and latent state sequences $\mathbf{s}_g$ (see Appendix A).

2.2. *Inference.* After the model fitting procedure described above, inference ultimately utilizes the gene-specific posterior probabilities of state sequences

(1)          $\Pr(\mathbf{s}_g = \mathbf{v} | \mathbf{x}_g) = \Pr(\mathbf{s}_g = \mathbf{v}) f(\mathbf{x}_g | \mathbf{s}_g = \mathbf{v}) / f(\mathbf{x}_g), \qquad \mathbf{v} \in \mathcal{S}$

(a gene-specific vector of the posterior probabilities of the 27 state sequences in the case of the CD8 T-cell data) and the gene-specific posterior distribution of the mean sequence

(2)          $f(\boldsymbol{\mu}_g | \mathbf{x}_g) = \sum_{\mathbf{v} \in \mathcal{S}} \Pr(\mathbf{s}_g = \mathbf{v} | \mathbf{x}_g) f(\boldsymbol{\mu}_g | \mathbf{x}_g, \mathbf{s}_g = \mathbf{v}),$

a mixture of the conditional posterior distribution over all possible state sequences. From (1) an optimal state sequence may be obtained either by separately maximizing the marginal posterior probability at each time point (MMP), or by maximizing the joint posterior probability over all time points (MJP). A gene for which the optimal state sequence entails some temporal changes is a candidate for a short list of interesting, temporally altered genes. The false discovery rate for such a list may be derived from the posterior probabilities themselves, as shown in Appendix B [see Benjamini and Hochberg (1995), Efron et al. (2001), Storey (2002, 2003), Dudoit, Shaffer and Boldrick (2003) and Newton et al. (2004) for further details about the false discovery rate]. From (2) a posterior summary statistic such as the posterior mean can be obtained to summarize $\boldsymbol{\mu}_g$.

## 3. Results.

3.1. *Analysis of the CD8 T-cell experiment.* We re-analyzed the CD8 T-cell experiment with the proposed hierarchical state space model to identify and to cluster genes involved in the development of memory CD8 T cells. Background correction, normalization and probe-set summaries were performed on the Affymetrix chip data using RMA from the Bioconductor *affy* package [Irizarry et al. (2003)]. Expression values were obtained on the log base 2 scale. Here data were analyzed using both a first-order hierarchical state space model (first-order HST) and a full-order hierarchical state space model (full-order HST). MMP and MJP were each applied to the gene-specific posterior probabilities of state sequences separately to obtain the gene-specific optimal state sequence.

Plots of the estimated population distributions for successive changes under the upregulation state and the downregulation state at each time point after the first time point are shown in Figure 2. Here dashed (solid) lines represent the estimated population distribution for successive changes under the upregulation (downregulation) state. The first-order HST model and the full-order HST model provide similar population distribution estimates.

Table 2 presents the estimated initial state probability vector and the state transition matrices of the discrete Markov process from the first-order HST model, suggesting a high correlation across time. In detail, compared to genes with no change at the previous time point, genes with change at the previous time point (state ="−" or "+") tend to have a much higher probability of change at the current time point. This pattern indicates that utilizing the sequential information has the potential of increasing the sensitivity of detecting TDE genes. The increase in sensitivity is also demonstrated in a simulation study following the analysis of the CD8 T-cell data. In addition, most genes that change in the first period (Naïve to d8) tend to have an opposite pattern in the second period (d8 to d15) where $\Pr(s_{g3} = "−"|s_{g2} = "+") = 0.68$ and $\Pr(s_{g3} = "+"|s_{g2} = "−") = 0.72$. This pattern indicates that utilizing the sequential information from the population has the potential of reducing the misclassification rate.

Table 3 shows the number of genes identified as TDE genes using different models and different optimality criteria (MMP and MJP). These four approaches yield similar numbers. More genes are identified as TDE in the earlier period than in the later period. Given the same model, conducting inference using the MMP results in more genes identified as TDE genes compared to conducting inference using the MJP. Given the same optimality criteria, more genes were identified as TDE genes under the full-order HST model than under the first-order HST model.

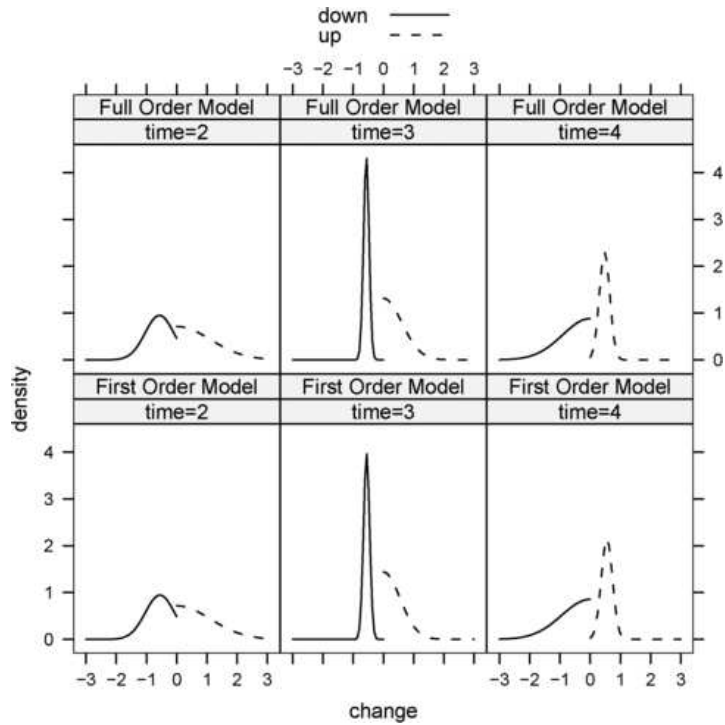

FIG. 2. *Estimated population distributions for successive changes under the upregulation and the downregulation state over time. Here dashed (solid) lines represent the estimated population distribution for successive changes under the upregulation (downregulation) state.*

TABLE 2
*Estimated parameters of the first-order Markov process in the state level*

| Naïve to d8 | | | Transition from d8 to d15 | | | | Transition from d15 to Imm | | | |
|---|---|---|---|---|---|---|---|---|---|---|
| + | − | = | d8\d15 | + | − | = | d15\Imm | + | − | = |
| 0.04 | 0.08 | 0.88 | + | 0.09 | 0.68 | 0.23 | + | 0.33 | 0.16 | 0.51 |
| | | | − | 0.72 | 0.00 | 0.28 | − | 0.00 | 0.18 | 0.82 |
| | | | = | 0.00 | 0.00 | 1.00 | = | 0.00 | 0.00 | 1.00 |

Clustering results from the first-order HST model are shown in Figure 3. The first-order and the full-order HST models produce very similar results

TABLE 3

*Number of genes identified as TDE genes between each two adjacent time points under different models (first-order and full-order HST models) and different optimality criteria (MMP and MJP)*

| Model | Criteria | d8 | d15 | Imm | Across time series |
|---|---|---|---|---|---|
| First-order model | MMP | 1318 | 951 | 328 | 1318 |
| | MJP | 1244 | 951 | 351 | 1244 |
| Full-order model | MMP | 1310 | 803 | 450 | 1342 |
| | MJP | 1215 | 815 | 444 | 1245 |

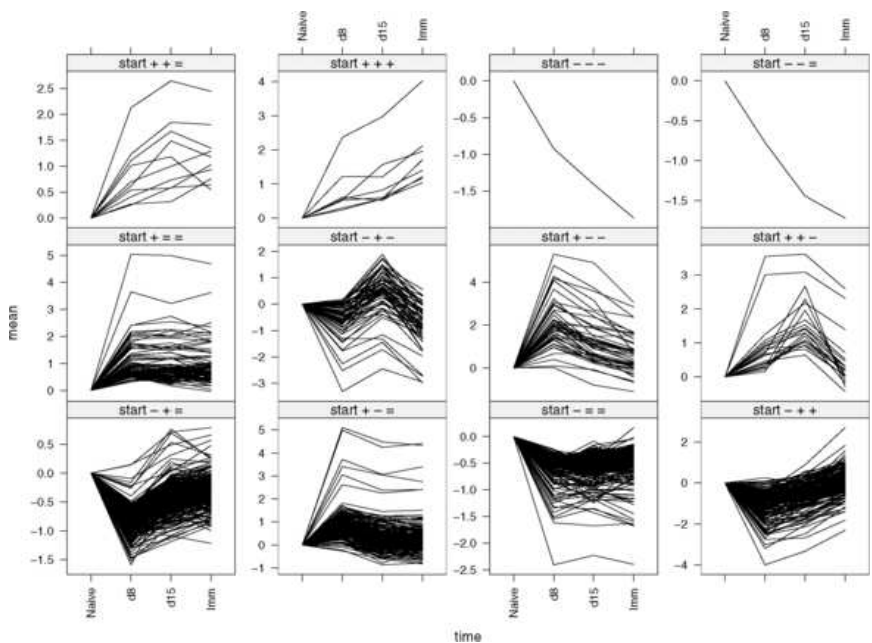

FIG. 3. *Clustering results from the first-order HST model. Here x-axis represents time and y-axis represents centered posterior mean so that they all start from 0.*

in the state level. Using the MMP, for example, the first-order HST model identifies 13 patterns and the full-order HST model identifies 15 patterns with a total of 11 patterns in common. With respect to all the genes, about 97.7% of the genes share the same optimal state series in these two models. Among genes identified as TDE in either of these models (1318 genes for first-order HST model and 1342 genes for full-order HST model), 1308 genes are in common. For these 1308 genes, about 81.5% share the same optimal state series, about 98.3% share the same optimal state at time point 2, about 87.7% share the same optimal state at time point 3, and about 92.0% share the same optimal state at time point 4, indicating a good consistency between the first-order and the full-order HST models.

We compared our results from the first-order HST model with the MMP criteria to the results obtained from the original analysis using the K-means clustering algorithm [Kaech et al. (2002a)]. Both methods identify clusters "start, $+$, $-$, $-$," "start, $-$, $+$, $+$," and "$+$, $=$, $=$." Information on these clusters is shown below:

**Cluster "start, $+$, $-$, $-$":** Upregulated at day 8 and gradually downregulated at day 15 and memory stage. This cluster contains 46 genes including: genes related to T cell effector functions such as GZMA, GZMB and GZMK; genes related to cell adhesion and migration such as CCR2; genes related to membrane proteins such as KLRG1.

**Cluster "start, $-$, $+$, $+$":** Downregulated at day 8 and gradually upregulated at day 15 and memory stage. This cluster contains 192 genes including genes related to T cell signal transduction such as IL7R; genes related to apoptosis/survival such as BCL2; genes related to cell adhesion and migration such as CXCR4 and CD62L.

**Cluster "start, $+$, $=$, $=$":** Upregulated at day 8 with no change over the following time points. This cluster contains 282 genes including genes related to cell adhesion and migration such as CD44 and genes related to membrane proteins such as LY6A.

Our method also identifies new clusters including the following:

**Cluster "start, $+$, $+$, $+$":** Continuously upregulated at all time points after the first time point and involved in the differentiation of memory CD8 T cells, this cluster of genes can be used to differentiate naïve, effector and memory CD8 T cells. This cluster contains 8 genes including genes related to T cell effector functions such as IFNg and genes related to cell adhesion and migration such as CXCR3.

**Cluster "start, $+$, $-$, $=$":** Upregulated at day 8, downregulated at day 15, and no change at memory stage. This cluster contains 282 genes including genes related to T cell effector functions such as FASL and genes related to cell adhesion and migration such as CCR5.

In addition, no important biomarkers fall into the cluster with pattern "start, $-$, $+$, $-$." Examining genes that fall into this cluster, we found that few genes have function or pathway related to the immune response, according to currently available annotation information from Gene Ontology [Vinayagam et al. (2004)]. These observed changes may be just due to random variation. Compared to previous clusters obtained by the K-means clustering, our model identified more clusters with clearer and more meaningful patterns. Based on clusters produced from our model, the investigators may more readily identify patterns involved in the CD8 T cell differentiation and focus on genes with patterns similar to important biomarkers.

3.2. *Simulation.* In addition to the analysis of the CD8 T-cell data, a simulation study was carried out to evaluate the performance of the proposed model. We generated 100 data sets each with 4,000 genes and four replicates at each of 4 time points. Data were produced from the first-order HST model in the order of state series, mean gene expression series and gene expression levels. We simulated the data using parameters estimated from the CD8 T-cell data to mimic the real scenario. First, we draw state series independently from a first-order Markov process (see Table 2). Then, we draw mean gene expression levels at the first time point independently from a Normal distribution and draw relative changes under the upregulation (downregulation) state independently from the 0-left-truncated (0-right-truncated) Normal distribution (see Figure 2). In turn, observed gene expression levels were independently produced from a Normal distribution conditional on the mean gene expression levels. Simulated data sets were analyzed using the following three methods: (I) a first-order HST model, (II) a zero-order HST model (independent HST model), and (III) a pairwise analysis method. Here the pairwise analysis was done by comparing each of the three successive changes in mean gene expression levels separately, ignoring changes at the other periods. Model parameters were estimated using the EM algorithm and the MMP criterion was applied to obtain the gene-specific optimal state sequence. Performance of these methods was assessed in both the discrete state level and the continuous mean level.

In the state level, we evaluated these three methods by comparing specificity, sensitivity, the false discovery rate, and the misclassification rate of the states at each time point along with the misclassification rate of the state series. Here the misclassification rate of the states (state series) is the proportion of genes whose optimal state (state series) does not match the true state (state series). Define the state of no change ("=') as the temporally nondifferentially expressed (TNDE) state and the other two states ("+" or "−") as the temporally differentially expressed (TDE) state. At each time point after the first time point, the sensitivity is the proportion of TDE states that are correctly identified as TDE, the specificity is the proportion of TNDE states that are correctly identified as TNDE, and the false discovery rate is the proportion of false TDE states among those identified as TDE.

Table 4 and Figure 4 summarize the simulation results in the state level. The first-order HST model has a lower misclassification rate than the independent HST model, which in turn has a lower misclassification rate than the pairwise method. Noticeably, the first-order HST model reduces the misclassification rate of state series by 32% (49%) compared to the independent

TABLE 4
*Summary of simulation results in the state level. Here I represents the first-order HST model, II represents the independent HST model, and III represents the pairwise method. MR is the misclassification rate of states at each time point and SMR is the misclassification rate of state series*

|  | Method | Time 1 to Time 2 | Time 2 to Time 3 | Time 3 to Time 4 |
|---|---|---|---|---|
| FDR | I | 0.158 (0.026) | 0.091 (0.020) | 0.119 (0.036) |
|  | II | 0.197 (0.031) | 0.066 (0.022) | 0.164 (0.047) |
|  | III | 0.191 (0.027) | 0.397 (0.038) | 0.506 (0.047) |
| Sensitivity | I | 0.760 (0.024) | 0.678 (0.030) | 0.683 (0.043) |
|  | II | 0.639 (0.024) | 0.407 (0.025) | 0.581 (0.044) |
|  | III | 0.667 (0.024) | 0.556 (0.027) | 0.647 (0.042) |
| Specificity | I | 0.980 (0.004) | 0.993 (0.002) | 0.997 (0.001) |
|  | II | 0.979 (0.004) | 0.997 (0.001) | 0.996 (0.002) |
|  | III | 0.978 (0.004) | 0.964 (0.006) | 0.976 (0.005) |
| MR | I | 0.047 (0.003) | 0.035 (0.003) | 0.014 (0.002) |
|  | II | 0.063 (0.004) | 0.056 (0.002) | 0.019 (0.002) |
|  | III | 0.059 (0.004) | 0.073 (0.005) | 0.035 (0.004) |
| SMR | I | 0.069 (0.004) | 0.069 (0.004) | 0.069 (0.004) |
|  | II | 0.102 (0.004) | 0.102 (0.004) | 0.102 (0.004) |
|  | III | 0.136 (0.008) | 0.136 (0.008) | 0.136 (0.008) |

HST model (the pairwise analysis). In addition, the first-order HST model increases the sensitivity by as much as 22% compared to the pairwise analysis. Compared to the pairwise analysis, the hierarchical state space model helps to improve the specificity, reduce the false discovery rate, and reduce the misclassification rate. Using a first-order HMM structure in the state level in the first-order HST model helps to reduce the misclassification rate and improve the sensitivity. In summary, these results indicate that the first-order HST model outperforms the other two methods in the state level. In addition, we investigate the performance of our model in the mean level. Using the mean from the posterior distribution of the mean gene expression level obtained from (2) in the first-order HST model as the summary, we find that posterior mean gene expression level reduces the variance and mean square error by 53% (from 0.019 to 0.009) compared to the sample mean gene expression level.

**4. Discussion.** In this article we analyzed a short time course microarray experiment to investigate the differentiation of memory CD8 T cells. To understand the differentiation of memory CD8 T cells, we need to detect

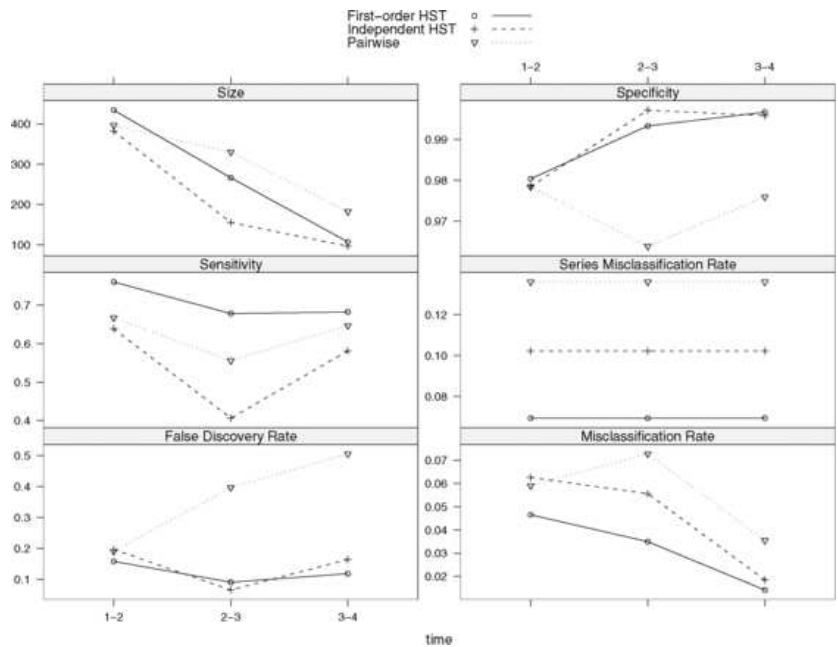

Fig. 4. *Summary of simulation results in the state level.*

TDE genes, identify the direction of successive changes over time, and characterize the magnitude of successive changes over time. Considering this, we develop a hierarchical state space model. To improve inference, a HMM structure is exploited to utilize the correlation of states across time and an empirical Bayes framework is utilized to borrow information from the large gene population. Results from the CD8 T-cell data indicate a strong correlation over time in the state level and an empirical Bayes analysis helps us to capture this theme from the large gene population (see Table 2 for details). Results from a simulation study show that incorporating this correlation may help to increase the sensitivity of detecting TDE genes and to reduce the misclassification rate (see Table 4 for details).

To analyze the CD8 T-cell experiment, we developed a specific hierarchical state space model. The proposed model can also be easily applied to other short time course microarray experiments which share a similar structure to the CD8 T-cell experiment (one biological condition and independent sampling). In addition, the specifications of the proposed model could be adjusted to suit different situations. For example, in the observation level, we could allow the sampling variance to vary over time instead of being time-independent. In the mean level, we could use a Gamma distribution, instead of a truncated Normal distribution, to model the successive changes over time. More details can be found in Wu (2007).

The proposed model can be extended in a straightforward manner to experiments with a longitudinal sampling design where subjects involved in experiments have repeated observations over time. For example, in a study of monkeys infected with simian immunodeficiency virus, the same group of monkeys were observed over time after infection [see Wu (2007) for more details]. The proposed hierarchical state space model could also be extended to time course microarray experiments with long time series. However, since the number of possible state series increases exponentially with the number of the time points ($3^{T-1}$), inference using the EM algorithm would be computationally intensive. Updating the posterior distribution using Monte Carlo simulation with a sequential strategy [Liu and Chen (1998)] could be a promising approach for analyzing time course microarray data with long time series.

Analysis of the CD8 T-cell data using the proposed hierarchical state space model produces biologically meaningful results. With explicit optimal state series, genes can be easily grouped and studied, either marginally or jointly. Important temporal patterns involved in biological processes can be quickly identified and genes that have a pattern similar to known important biomarkers can be easily extracted for further investigation. These results may also be used for further high level analysis, such as network and pathway reconstruction. Analysis of the immune response suggests that multiple pathways are involved in the differentiation of memory CD8 T cells, including cell adhesion/migration, signal transduction and effector response. Here we want to emphasize that this analysis is conducted at the genetic level. Providing rich insight into biological processes, these results also require careful interpretation to answer questions of interest. For example, a high level of gene expression does not necessarily yield a high level of its functional protein. In immune response experiments, the gene expression levels for some genes related to antigen-killing functions such as GZMB and IFNG are higher in memory CD8 T cells than in naïve CD8 T cells. However, there is no corresponding high production of effector response-related proteins. This amazing strategy enables memory CD8 T cells to have a quick recall response without releasing improper cytotoxic proteins harming healthy cells [Veiga-Fernandes et al. (2000)]. It is desirable to combine information from other biological studies to obtain a complete understanding of the differentiation of memory CD8 T cells in response to a viral infection.

## APPENDIX A: MODEL FITTING

Under the gene-wise independence assumption, the complete log-likelihood is

$$\sum_{g=1}^{G} \log f(\mathbf{x}_g, \boldsymbol{\mu}_g, \mathbf{s}_g)$$

$$= \sum_{g=1}^{G} \log f(\mathbf{x}_g | \boldsymbol{\mu}_g) + \sum_{g=1}^{G} \sum_{\mathbf{v} \in \mathcal{S}} I[\mathbf{s}_g = \mathbf{v}][\log f(\boldsymbol{\mu}_g | \mathbf{s}_g = \mathbf{v}) + \log \Pr(\mathbf{v})].$$

One may question this gene-wise independence assumption. However, considering that parameter estimates in the proposed model are driven largely by the marginal distribution of the data, parameter estimates in a large-scale study like the microarray study ought to be reliable (though probably estimated less well than expected) in the presence of modest among-gene dependence. Treating $(\boldsymbol{\mu}_1, \dots, \boldsymbol{\mu}_G)$ and $(\mathbf{s}_1, \dots, \mathbf{s}_G)$ as missing data, we employ the EM algorithm to find the MLE iteratively. Here we use the proposed parametric first-order HST model to illustrate the fitting procedure. Given thousands of genes in the experiments, $\sigma^2$ can be reliably estimated using the usual unbiased estimator

$$\hat{\sigma}^2 = \frac{1}{G \times (\sum_{t=1}^{T} n_t - T)} \sum_{g=1}^{G} \sum_{t=1}^{T} \sum_{k=1}^{n_t} (x_{gtk} - \bar{x}_{gt})^2,$$

where $\bar{x}_{gt}$ is the sample mean of the gene expression level for gene $g$ at the $t$th time point.

In the expectation step, we can estimate $\Pr(\mathbf{s}_g = \mathbf{v} | \mathbf{x}_g)$ by

$$\widehat{\Pr}(\mathbf{s}_g = \mathbf{v} | \mathbf{x}_g) = \frac{\widehat{\Pr}(\mathbf{s}_g = \mathbf{v}) \hat{f}(\mathbf{x}_g | \mathbf{s}_g = \mathbf{v})}{\hat{f}(\mathbf{x}_g)}, \qquad \mathbf{v} \in \mathcal{S}.$$

From $\widehat{\Pr}(\mathbf{s}_g | \mathbf{x}_g)$, we can estimate $\Pr(\mathbf{s}_{gt} = i | \mathbf{x}_g)$ by $\sum_{\mathbf{v}_t = i} \widehat{\Pr}(\mathbf{s}_g = \mathbf{v} | \mathbf{x}_g)$ and $\Pr(\mathbf{s}_{g(t-1)} = i, \mathbf{s}_{gt} = j | \mathbf{x}_g)$ by $\sum_{\mathbf{v}_{t-1} = i, \mathbf{v}_t = j} \widehat{\Pr}(\mathbf{s}_g = \mathbf{v} | \mathbf{x}_g)$ for $i, j = +, -, =$ and $t = 2, \dots, T$. In the maximization step, the initial state probabilities and the state transition matrices can be updated by

$$\hat{\pi}_i = \widehat{\Pr}(S_t = i) = \frac{1}{G} \sum_{g=1}^{G} \widehat{\Pr}(s_{gt} = i | \mathbf{x}_g),$$

$$\widehat{\prod}_{ij}^{t} = \widehat{\Pr}(S_{t-1} = i, S_t = j) = \frac{1}{G} \sum_{g=1}^{G} \widehat{\Pr}(s_{g(t-1)} = i, s_{gt} = j | \mathbf{x}_g)$$

and model parameters in the mean level can be updated by maximizing

$$\prod_{g=1}^{G} \sum_{i=+,-,=} \widehat{\Pr}(s_{gt} = i | \mathbf{x}_g) f(x_{g(t-1)}, x_{gt} | s_{gt} = i).$$

Iterate between the expectation step and the maximization step until convergence.

To focus on the successive changes over time, a simple center transformation, $\tilde{x}_{gtk} = x_{gtk} - \bar{x}_{g1}$, can be used to simplify the proposed model. After the center transformation, the proposed model can be rewritten as

$$\tilde{x}_{gtk} = \tilde{\mu}_{gt} + \epsilon_{gtk},$$

$$\tilde{\mu}_{g1} \sim N(0, \sigma^2/n_1),$$

$$\tilde{\mu}_{gt} = \tilde{\mu}_{g(t-1)} + \delta_{gt},$$

with other parts of the proposed hierarchical state space model unchanged. Here $\tilde{x}_{gtk}$ represents the gene expression level after the center transformation and $\tilde{\mu}_{gt}$ represents the latent mean after the center transformation. Centering may induce some dependence as discussed in Dahl and Newton (2007).

## APPENDIX B: CONDITIONAL FDR

Denoting by $C_t$ the set of genes identified as TDE at each time point and denoting by $C$ the combined set of genes identified as TDE, the expected FDR at each time point can be estimated by

$$\widehat{\mathrm{FDR}}_t = \frac{\sum_{g=1}^{G} \widehat{\mathrm{Pr}}(s_{gt} = ``="|\mathbf{x}_g) I(g \in C_t)}{\sum_{g=1}^{G} I(g \in C_t)},$$

the average of the conditional false discovery rate at time $t$ for genes in $C_t$. Similarly, the overall expected FDR can be estimated by

$$\widehat{\mathrm{FDR}} = \frac{\sum_{g=1}^{G} \widehat{\mathrm{Pr}}(\text{no change over time}|\mathbf{x}_g) I(g \in C)}{\sum_{g=1}^{G} I(g \in C)},$$

the average of the conditional false discovery rate over all time points for genes in $C$ [see Newton et al. (2004) for further elaboration on this point].

## REFERENCES

ALTER, O., BROWN, P. O. and BOTSTEIN, D. (2000). Singular value decomposition for genome-wide expression data processing and modeling. *Proc. Natl. Acad. Sci. USA* **97** 10101–10106.

BAR-JOSEPH, Z., GERBER, G., GIFFORD, D. K., JAAKKOLA, T. S. and SIMON, I. (2002). A new approach to analyzing gene expression time series data. In *Annual Conference on Research in Computational Molecular Biology Proceedings of the Sixth Annual International Conference on Computational Biology* 39–48. ACM Press, New York.

BENJAMINI, Y. and HOCHBERG, Y. (1995). Controlling the false discovery rate: A practical and powerful approach to multiple testing. *J. Roy. Statist. Soc. Ser. B* **57** 289–300. MR1325392

DAHL, D. B. and NEWTON, M. A. (2007). Multiple hypothesis testing by clustering treatment effects. *J. Amer. Statist. Assoc* **102** 517–526. MR2325114

DEMPSTER, A. P., LAIRD, N. M. and RUBIN, D. B. (1977). Maximum likelihood from incomplete data via the EM algorithm. *J. Roy. Statist. Soc. Ser. B* **39** 1–38. MR0501537

DUDOIT, S., SHAFFER, J. P. and BOLDRICK, J. C. (2003). Multiple hypothesis testing in microarray experiments. *Statist. Sci.* **18** 71–103. MR1997066

EFRON, B., TIBSHIRANI, R., STOREY, J. D. and TUSHER, V. (2001). Empirical Bayes analysis of a microarray experiment. *J. Amer. Statist. Assoc.* **96** 1151–1160. MR1946571

EISEN, M. B., SPELLMAN, P. T., BROWN, P. O. and BOTSTEIN, D. (1998). Cluster analysis and display of genome-wide expression patterns. *Proc. Natl. Acad. Sci. USA* **95** 14863–14868.

ERNST, J., NAU, G. J. and BAR-JOSEPH, Z. (2005). Clustering short time series gene expression data. *Bioinformatics* **21** i159–i168.

GOLLUB, J., BALL, C. A., BINKLEY, G., DEMETER, J., FINKELSTEIN, D. B., HEBERT, J. M., HERNANDEZ-BOUSSARD, T., JIN, H., KALOPER, M., MATESE, J. C., SCHROEDER, M., BROWN, P. O., BOTSTEIN, D. and SHERLOCK, G. (2003). The Stanford microarray database: Data access and quality assessment tools. *Nucleic Acids Research* **31** 94–96.

HEARD, N. A., HOLMES, C. C. and STEPHENS, D. A. (2006). A quantitative study of gene regulation involved in the immune response of anopheline mosquitoes: An application of Bayesian hierarchical clustering of curves. *J. Amer. Statist. Assoc.* **101** 18–29. MR2252430

HONG, F. and LI, H. (2006). Functional hierarchical models for identifying genes with different time-course expression profiles. *Biometrics* **62** 534–544. MR2236847

IRIZARRY, R. A., HOBBS, B., COLLIN, F., BEAZER-BARCLAY, Y. D., ANTONELLIS, K. J., SCHERF, U. and SPEED, T. P. (2003). Exploration, normalization and summaries of high density oligonucleotide array probe level data. *Biostatistics* **4** 249–264.

KAECH, S. M., HEMBY, S., KERSH, E. and AHMED, R. (2002a). Molecular and functional profiling of memory CD8 T cell differentiation. *Cell* **111** 837–851.

KAECH, S. M., WHERRY, E. J. and AHMED, R. (2002b). Effector and memory T cell differentiation: Implications for vaccine development. *Nature Review Immunology* **2** 251–262.

KENDZIORSKI, C. M., NEWTON, M. A., LAN, H. and GOULD, M. N. (2003). On parametric empirical Bayes methods for comparing multiple groups using replicated gene expression profiles. *Statistics in Medicine* **22** 3899–3914.

KLEVECZ, R. R. (2000). Dynamic architecture of the yeast cell cycle uncovered by wavelet decomposition of expression microarray data. *Functional and Integrative Genomics* **1** 186–192.

LIU, J. S. and CHEN, R. (1998). Sequential Monte Carlo methods for dynamic systems. *J. Amer. Statist. Assoc.* **93** 1032–1044. MR1649198

NEWTON, M. A., KENDZIORSKI, C. M., RICHMOND, C. S., BLATTNER, F. R. and TSUI, K. W. (2001). On differential variability of expression ratios: Improving statistical inference about gene expression changes from microarray data. *J. Comput. Biol.* **8** 37–52.

NEWTON, M. A., NOUEIRY, A., SARKAR, D. and AHLQUIST, P. (2004). Detecting differential gene expression with a semiparametric hierarchical mixture method. *Biostatistics* **5** 155–176.

PARK, T., YI, S.-G., LEE, S., LEE, S. Y., YOO, D.-H., AHN, J.-I. and LEE, Y.-S. (2003). Statistical tests for identifying differentially expressed genes in time-course microarray experiments. *Bioinformatics* **19** 694–703.

RAMONI, M., SEBASTIANI, P. and COHEN, P. (2002). Bayesian clustering by dynamics. *Machine Learning* **47** 91–121.

Schliep, A., Schönhuth, A. and Steinhoff, C. (2003). Using hidden Markov models to analyze gene expression time course data. *Bioinformatics* **19** i255–i263.

Schliep, A., Steinhoff, C. and Schönhuth, A. (2004). Robust inference of groups in gene expression time-courses using mixtures of HMMs. *Bioinformatics* **20** i283–i289.

Spellman, P. T., Sherlock, G., Zhang, M. Q., Iyer, V. R., Anders, K., Eisen, M. B., Brown, P. O., Botstein, D. and Futcher, B. (1998). Comprehensive identification of cell cycle-regulated genes of the yeast *Saccharomyces cerevisiae* by microarray hybridization. *Molecular Biology of the Cell* **9** 3273–3297.

Storey, J. D. (2002). A direct approach to false discovery rates. *J. Roy. Statist. Soc. Ser. B* **64** 479–498. MR1924302

Storey, J. D. (2003). The positive false discovery rate: A Bayesian interpretation and the q-value. *Ann. Statist.* **31** 2013–2035. MR2036398

Storey, J. D., Xiao, W., Leek, J. T., Tompkins, R. G. and Davis, R. W. (2005). Significance analysis of time course microarray experiments. *Proc. Natl. Acad. Sci. USA* **102** 12837–12842.

Tai, Y. C. and Speed, T. P. (2006). A multivariate empirical Bayes statistic for replicated microarray time course data. *Ann Statist.* **34** 2387–2412. MR2291504

Tamayo, P., Slonim, D., Mesirov, J., Zhu, Q., Kitareewan, S., Dmitrovsky, E., Lander, E. S. and Golub, T. R. (1999). Interpreting patterns of gene expression with self-organizing maps: Methods and application to hematopoietic differentiation. *Proc. Natl. Acad. Sci. USA* **96** 2907–2912.

Tavazoie, S., Hughes, J. D., Campbell, M. J., Cho, R. J. and Church, G. M. (1999). Systematic determination of genetic network architecture. *Nature Genetics* **22** 281–285.

Veiga-Fernandes, H., Walter, U., Bourgeois, C., McLean, A. and Rocha, B. (2000). Response of naïve and memory CD8+ T cells to antigen stimulation in vivo. *Nature Immunology* **1** 47–53.

Vinayagam, A., Pugalenthi, G., Rajesh, R. and Sowdhamini, R. (2004). SDBASE: A consortium of native and modelled disulphide bonds in proteins. *Nucleic Acids Research* **32** 200–202.

Wall, M. E., Dyck, P. A. and Brettin, T. S. (2001). SVDMAN—singular value decomposition analysis of microarray data. *Bioinformatics* **17** 566–568.

Wu, H. (2007). Hierarchical analysis of microarray experiments with applications to the study of CD8 T cell immune respones. Ph.D. thesis, Emory Univ.

Yuan, M. and Kendziorski, C. (2006). Hidden Markov models for microarray time course data in multiple biological conditions (with discussion). *J. Amer. Statist. Assoc.* **101** 1323–1340.

Zhou, C. and Wakefield, J. (2006). A Bayesian mixture model for partitioning gene expression data. *Biometrics* **62** 515–525. MR2236834

H. Wu
Department of Biostatistics
Rollins School of Public Health
Emory University
1518 Clifton Rd.
Atlanta, Georgia 30322
USA
E-mail: haiyanwood@msn.com

M. Yuan
School of Industrial
  and Systems Engineering
Georgia Institute of Technology
Atlanta, Georgia 30332
USA
E-mail: myuan@isye.gatech.edu

S. M. Kaech
School of Medicine
Yale University
New Haven, Connecticut 06520
USA
E-mail: susan.kaech@yale.edu

M. E. Halloran
Department of Biostatistics
Fred Hutchinson Cancer
    Research Center
and
School of Public Health
    and Community Medicine
University of Washington
Seattle, Washington 98195
USA
E-mail: betz@u.washington.edu